# Minimizing the Time of Detection of Large (Probably) Prime Numbers

**Dragan Vidakovic**
assistant Gimnazija
Ivanjica, Serbia

**Dusko Parezanovic**
assistant Gimnazija
Ivanjica, Serbia

**Zoran Vucetic**
assistant Gimnazija
Ivanjica, Serbia

## ABSTRACT

In this paper we present the experimental results that more clearly than any theory suggest an answer to the question: when in detection of large (probably) prime numbers to apply, a very resource demanding, Miller-Rabin algorithm. Or, to put it another way, when the dividing by first several tens of prime numbers should be replaced by primality testing?

As an innovation, the procedure above will be supplemented by considering the use of the well-known Goldbach's conjecture in the solving of this and some other important questions about the RSA cryptosystem, always guided by the motto "do not harm" – neither the security nor the time spent.

## 1. INTRODUCTION

In asymmetric schemes [1] of protecting the confidentiality and integrity of data there is a need for large prime numbers. For some tasks required number of bits now exceeds 15,000, and it is still just a passing figure in the endless game of those who protect data and those who attack them. It is therefore quite clear that the time spent on detection of large prime numbers must be as short as possible.

It would be best to check the divisibility of number n with all prime numbers less than or equal to sqrt(n). However, with so many bits it's not realistic. Therefore, the number which is tested to primality is previously divided by several tens of first prime numbers and then, if it is not divisible by any of these numbers, it is left to Miller-Rabin algorithm [1].

It is a very difficult task to find theoretically an optimum ratio of time required for dividing the number and testing by Miller-Rabin algorithm. Perhaps a redundant task as well in terms of our needs, since in practical tasks such as ours, we have no reason to pretend that computers do not exist, that the experimentally obtained, very useful results are less valuable than the values obtained theoretically.

As a useful tool for our task (minimizing the time required for detection of prime numbers) we see the Goldbach conjecture [2], which states that every (large for us) even number is the sum of two prime numbers. It may forever remain a conjecture, or one day some talented mathematician may write a book of hundreds of pages that will prove its truth, or some computer may find the number for which it is not valid, and with that break the conjecture.

For those of us looking for large prime numbers none of these three matters. We will, in any case, generate a random large even number 2n, of, say, 1024 bits, and detect a much smaller random prime number of, say, 128 or 256 bits, which is negligible in terms of time, and then verify that the difference of those two numbers is a prime number. If so, we have a large prime number. If not, we will repeat the procedure or we will use this difference to generate the prime number closest to it by a combination of dividing by first prime numbers and Miller-Rabin algorithm. Experimentally we will ensure that the above procedures may also result in saving the time required to detect a large prime number.

## 2. WHY WE NEED PRIME NUMBERS

The public key cryptography-PK [1][3], a major breakthrough in the field of data secrecy and integrity protection, is mostly based on the assurance which has never been mathematically proved that some mathematical problems are difficult to solve. The two of them are particularly prominent and used a lot.

Since we opted for RSA [1][3] mechanism we will point to one of them. The multiplication of two large prime numbers is a one-way hash function [1], which means that we can easily get their multiplication result. However, factorization of that multiplication result with the aim of getting the prime factors (factoring), turned out to be very difficult. This problem of identifying private key d in the Public Key cryptography (PK), if we know the public key and if it is the pair (n,e) are two equivalent problems [1][3]. Certainly, there are many other PK schemes, asymmetrical algorithms, apart from RSA. They are based on the same problem which is difficult to solve in practice if the number of digits is large enough, and by means of these schemes a one-way function with “trap door” is created.

By technological development and progress in the field of algorithms for whole number factorization, the need for larger and larger prime numbers has been demonstrated. This means that their multiplication result will consist of more and more digits. The competition between those who attack unprotected data and those who protect them using RSA mechanism requires creation of the faster operations for dealing with large numbers. The new arithmetic requires more efficient codes for addition, subtraction, multiplication and division of large numbers and what is particularly significant is to solve modular exponentiation in the most efficient way possible [4].

This makes sense only if special attention is paid to the creation of one's large (probable) prime numbers, since the use of such numbers available on the Internet or in any other way is not in accordance with the very aim of data protection. Since the process of large prime generation requires a lot of time and computer resources [1][4], it is of particular interest to us to find a way to avoid the application of the primality testing algorithm to the number as much as possible.

## 3. EXPERIMENTAL RESULTS

In order to avoid unnecessary applications of Miller-Rabin algorithm to the number in question, we resort to trial division by a few initial prime numbers, since such a division take less time.

How far we should go with such a division is the that we are trying to answer in this paper? For the theory of the matter is fully resolved. However, that in practice we do not have much use.

The trial division takes less time then exponentiation [1][4], but it would certainly be wrong to conclude that we should divide the number as long as possible. It is very difficult to determine the real relation between the two, since everything depends on the number we start with and odd numbers we examine so as to generate a probable prime.

Therefore, we present two solutions that are probably irrelevant to theorists, but it is very useful to people who have spent many nights to produce large (probably) prime numbers using its own software [4].

### 3.1 Dividing by First Several Tens of Prime Numbers

In this paragraph we show the results of detection of prime numbers of 513, 1024 and 1500 bits, namely: without dividing by prime numbers, dividing by first 10, 20, 30,…, 100 prime numbers.

***Example 1***

If we start with number c with ones in places: c[512]:=1; c[255]:=1; c[200]:=1;c[127]:=1; c[100]:=1; c[50]:=1; c[10]:=1; c[9]:=1; c[8]:=1; c[7]:=1; c[2]:=1; c[1]:=1; c[0]:=1; by dividing and testing we intend to detect first prime number with ones in places: c[512]:=1; c[255]:=1;

c[200]:=1;c[127]:=1; c[100]:=1; c[50]:=1; c[11]:=1; c[9]:=1; c[6]:=1; c[3]:=1; c[2]:=1; c[1]:=1; c[0]:=1; as a result we have the following table:

**TABLE 1**. **The Timing of Detection of a Prime Numbers**

| | | | | | | | | | | |
|---|---|---|---|---|---|---|---|---|---|---|
| **a** | 353 | 110 | 91 | 81 | 73 | 72 | 67 | 66 | 65 | 65 |
| **b** | 0 | 10 | 20 | 30 | 40 | 50 | 60 | 70 | 80 | 90 |
| **c** | 1455” | 466” | 398” | 361” | 337” | 342” | 329” | 330” | 334” | 337” |
| **d** | 1455” | 453” | 375” | 334” | 301” | 297” | 276” | 272” | 268” | 268” |
| **e** | 0” | 13” | 23” | 27” | 36” | 45” | 53” | 58” | 66” | 69” |
| **a** | 63 | 62 | 61 | 61 | 61 | 60 | 58 | 57 | 57 | 56 |
| **b** | 100 | 110 | 120 | 130 | 140 | 150 | 160 | 170 | 180 | 200 |
| **c** | 326” | 328” | 331” | 337” | 343” | 345” | 343” | 344” | 353” | 358” |
| **d** | 260” | 255” | 251” | 251” | 251” | 247” | 239” | 235” | 235” | 231” |
| **e** | 66” | 73” | 80” | 86” | 92” | 98” | 104” | 109” | 118” | 127” |

Where the following row labels are valid:

**a-** number of passes through the Miller-Rabin (MR) algorithm

**b-** number of first prime numbers by which we divide the number tested before passing through the MR algorithm

**c-** the total time needed for detection of a prime number

**d-** time spent on the MR algorithm

**e-** time spent on the division by first prime numbers

(all times are expressed in seconds).

It is clear from the table that the search for prime numbers, without dividing by first prime numbers, is not an option. This is an unnecessary waste of time. Dividing by first ten prime numbers would be a minimum. Dividing by 60, 70, ... would be a good choice. The choice of tens of numbers more or less could make little savings or a small loss of time and would not significantly affect the quality of our task.

***Example 2***

If we start with number c with ones in places: c[1023]:=1; c[767]:=1; c[512]:=1; c[255]:=1; c[127]:=1; c[100]:=1; c[50]:=1; c[10]:=1; c[9]:=1; c[8]:=1; c[7]:=1; c[2]:=1; c[1]:=1; c[0]:=1; by dividing and testing we intend to detect first prime number with ones in places: c[1023]:=1; c[767]:=1; c[512]:=1; c[255]:=1; c[127]:=1; c[100]:=1; c[50]:=1; c[11]:=1;

c[10]:=1; c[4]:=1; c[2]:=1; c[1]:=1; c[0]:=1; as a result we have the following table:

**TABLE 2. The Timing of Detection of a Prime Numbers**

| | | | | | | | | |
|---|---|---|---|---|---|---|---|---|
| **a** | 584 | 178 | 129 | 115 | 111 | 107 | 105 | 101 |
| **b** | 0 | 10 | 30 | 50 | 60 | 70 | 80 | 100 |
| **c** | 18144” | 5583” | 4251” | 3872” | 3778” | 3711” | 3734” | 3792” |
| **d** | 18144” | 5518” | 3999” | 3565” | 3441” | 3317” | 3255” | 3131” |
| **e** | 0” | 65” | 252” | 307” | 337” | 394” | 479” | 661” |

A minimum below which we should not go in generating 1024-bit prime numbers is dividing by first ten numbers, which reduces the time of detection of a prime number (about) three times. Dividing by first 60 to a hundred numbers reduces the time of detection of a prime number (about) five times, so that these values may be a good choice.

***Example 3***

If we start with number c with ones in places: c[1499]:=1; c[1023]:=1; c[767]:=1; c[512]:=1; c[255]:=1; c[127]:=1; c[100]:=1; c[50]:=1; c[10]:=1; c[9]:=1; c[8"1; c[7]:=1; c[2]:=1; c[1]:=1; c[0]:=1; by dividing and testing we intend to detect first prime number with ones in places: c[1499]:=1; c[1023]:=1; c[767]:=1; c[512]:=1; c[255]:=1; c[127]:=1; c[100]:=1; c[50]:=1; c[10]:=1; c[9]:=1; c[8]:=1; c[7]:=1; c[6]:=1; c[5]:=1; c[3]:=1; c[2]:=1; c[1]:=1; c[0]:=1; as a result we have the following table:

**TABLE 3. The Timing of Detection of a Prime Numbers**

| | | | | | | | |
|---|---|---|---|---|---|---|---|
| **a** | 50 | 16 | 15 | 14 | 12 | 12 | 12 |
| **b** | 0 | 10 | 20 | 30 | 40 | 50 | 60 |
| **c** | 4805” | 1551” | 1458” | 1387” | 1175” | 1178” | 1256” |
| **d** | 4805” | 1538” | 1442” | 1345” | 1153” | 1153” | 1153” |
| **e** | 0” | 13” | 16” | 42” | 22” | 25” | 103” |

Similar conclusions as in the above examples we may draw in the case of generating a 1500-bit number. Dividing by 50-60 first prime numbers is a very good choice.

### 3.2 Using Goldbach Conecture

Goldbach set the conjecture that "every even number (2n) larger than four is the sum of two (odd) prime numbers." [2]

Our idea is to detect a prime number p less than n and then to examine whether the difference (2n-p) is a prime number. Given that we choose p to be less than n (in order to detect it faster) and 2n to be a large number, which, if 2n-p is a prime, gives us a large prime number, avoiding the search through the upper part (numbers larger than n) which is in terms of time far more demanding job than detection of prime numbers in the lower part (numbers less than n).

Of course all of this is possible if there are enough pairs with simple coordinates between all pairs of numbers (p, q), where p + q = 2n, p-prime and q-odd number.

**TABLE 4. Number of GC Pairs**

| **Even number** | $2^{20}$ | $2^{21}$ | $2^{25}$ | $2^{26}$ | $2^{27}$ |
|---|---|---|---|---|---|
| **Number of Pairs GC** | 4244 | 7492 | 83543 | 153881 | 283830 |
| **Number of pairs ($*_1$)** | 43458 | 82125 | 1078257 | 2064123 | 3958400 |
| **% (GC) in ($*_1$)** | 9.77% | 9.12% | 7.75% | 7.45% | 7.17% |

The Table 4. shows that Goldbach conjecture can be a useful tool in our task because there is a probability, though not large, of guessing the large prime number. A possible loss of time in detecting the prime number for the first coordinate is negligible because it is number less than n, and it is particularly negligible compared to the possibility to immediately detect the other simple coordinate- a large prime number. We can get more favorable result, if we consider the set (*2 ) = {(p,r)} for given number 2n, p ≤ n-1, r ≥ n + 1, where p is a prime number and r is an odd number from the set {6*k+1, 6*k -1} and p + r = 2n. In any case, it is clear that by this process we cannot increase the time of detecting a large prime number, while we can significantly reduce it using favourable conditions.

With our own software we conducted an experiment whose aim was to find all pairs (p, q) for given number 2n, p ≤ n-1, q ≥ n + 1, where p is a prime number and q is an odd number and p + q = 2n (*1). Then, among these pairs to find those in which the second coordinate is prime number (pairs of Goldbach conjecture (GC)) and to measure the time of finding a number of representations of number 2n which satisfy the Goldbach conjecture.

## 4. SOME FURTER OBSERVATION

If we consider the time of finding all GC pairs of some even number 2n, we can see that with the increase of number n, the time to find them significantly increases.

**TABLE 5. Time to Find a GC Pairs**

| Even number | $2^{24}$ | $2^{25}$ | $2^{26}$ | $2^{27}$ |
|---|---|---|---|---|
| Number of pairs GC | 45752 | 83543 | 153881 | 283830 |
| Time of finding the pairs GC | 56’ 42” | 2h 1’ 14” | 3h 29’ 48” | 7h 18’ 17” |

### 4.1 Point “Multiplication”

One of the most important operations for all applications of elliptic curves cryptosystems (ECC) [5] is scalar point multiplication. Scalar point multiplication consists of calculating the value of an integer multiplied by a point by doing a series of point doublings and additions until the product point is reached.

***Example***

For NIST recommended prime field $F_p$ for $p= 2^{192}- 2^{64} -1$, base point $P(X_G, Y_G)$ [6]:

$X_G$ = 0x 188da80eb03090f67cbf20eb43a118800f4ff0afd82ff1012

$Y_G$ = 0x 07192b95ffc8da78631011ed6b24cdd53f977a11e794811, and

k=100000000000000000000000000000000000000000000000000000
1000000000000000000000000000000000000000000000010000000000
000000000000000000000000000010001101011

Coordinates of the point Q(X, Y) =k* $P(X_G, Y_G)$ are [6]:

X: 0x a9355c37074c8195faa23d1d071997fe4ea7a2bdec781047

Y:0x 911a232aabeba33b5e8f29743f2837955cd5bf1f74aa9a24

Very similar: Even if we know $2n = 2^{27}$, it takes more than seven hours to find a number of its GC representations. If we know the number of GC representations it is impossible to guess the number whose representations those are, which resembles a Goldbach "multiplication of point", which ends the "similarities" with ECC because we cannot define the addition of points. But, the question of using this sort of Goldbach’s point transformation (from starting point to the point (2n, (*1),(GC))) remains open for some further research.

### 4.2 The Possibility to Applying to the RSA

The public key technique developed by Rivest, Shamir and Adelman is known as the RSA algorithm. The security of this approach is based on the fact that it can be relatively easy to multiply large primes together but almost impossible to factor the resulting products. RSA has became the algorithm that most people associate with the notion of public key cryptography. The technique literally produces public keys that are tied to specific private keys. If Alice has a copy of Bob's public key she can encrypt a message to him, and he uses his private key to decrypt it. RSA also allows the holder of a private key to encrypt data with it so that anyone with a copy of the public key can then decrypt it. While public decryption obviously doesn't provide secrecy, the technique does provide digital signature, which attest that a particular crypto transform was performed by the owner of a particular private key [1][3].

RSA keys consist of three special numeric values that are used in pairs to perform encryption or decryption. The public key value is generally a selected constant that is recommended to be either 3 or 65537. After choosing the public key we generate two large prime numbers P and Q. The private key value is derived from P, Q, and the public key value. The distributed public keying material includes the constant public key value and the modulus N, which is the product of P and Q. The modulus is used in both the encryption and decryption procedures when either the public or private key is used. The original primes P and Q are discarded [1][3].

***Key generation for the RSA encryption:***

Each entity creates an RSA public key and a corresponding private key [1].

**Algorithm**

Each entity A should do the following:

- Generate two large distinct random primes p and q, each roughly the same size
- Compute $n = pq$ i $\phi = (p - 1)(q - 1)$.
- Select a random integer e, $1 < e < \phi$, tako da je nzd(e, $\phi$) = 1
- Use the extended Euclidean algorithm [1]. (Algorithm 2.107) to compute the unique integer d, $1 < d < \phi$, such that $ed \equiv 1 \pmod{\phi}$.
- A's public key is (n, e); A's private key is d

***Example 1***

Let the message m [4]: "rat" or binary: m = (0)10100100110000101110100

Select the two 128 BD primes:

p:
1000000000000000000000000010000000000000000000000000000000000
0000000000000000100000000000000000000000000000000000000010101
011111.

q:
1000000000000000000000000010000000000000000000000000000000000
0000000000000000100000000000000000000000000000100000000000111
001011

We count

n = pq:
1000000000000000000000000010000000000000000000000000001000000
0000000000000001000000000000000000000000000100010000000011100
1010100000000100000000011100101011000000000000000000000000000
0100000000011100101010000000000000000000001010101111100110100
00101010101.

$\phi = (p - 1)(q - 1)$:
1000000000000000000000000010000000000000000000000000001000000
0000000000000001000000000000000000000000000100010000000011100
1010000000000100000000011100101001000000000000000000000000000
0100000000011100101000000000000000000000001010101110100110011
01000101100.

Public key e = 3 or binary 11.

Secret key d:
1010101010101010101010101011010101010101010101010101100000000
0000000000000001010101010101010101010101100000101010101111011
1000000000001010101011110111000010101010101010101010101010101
1010101011010000110101010101010101010101110001111110001000100
0101110011.

Encrypted message m:
1000100000111111010101011010010110000011000101100100111000000010
1000000.

***Example 2***

Let the message m [4]: “Ratne godine!”

binary:
m=(0)101001001100000101110100011011100110010100100000011001110
1101111011001000110100101101110011001010100100001

Encrypted message m:

1011111001011001110110101110101110011111011010011010100110000
0011011001000101010101000010100101001110001011010101101100111
1011000001001001100000100100110111110100100101110001001001010
0111101110011001100001000111000010101011010010101010010000010
1011110101.

Now we point out the two possible connection between RSA and the Goldbach conjecture.

#### *4.2.1 The First Possibility*

Only powerful computers can calculate (GC) numbers of 1024, 2048, or a larger number of bits. We have no reasons not to believe that $(*_1)$ and (GC) are greater and greater numbers and at the same time (probably) they are unique for a given number 2n. Even if various even numbers have the same representation it does not matter for us because we will create a table that will contain in each row for a given even number the hash value $(*_1)$ and (GC).

For large, probably prime numbers $p_1$ and $q_1$ we will calculate the number $2n = p_1 * q_1 + 1$.

For this number we will find $(*_1)$ and (GC) and their hash values h $(*_1)$ and h (GC). The procedure will be repeated k times and the table of k rows (each of which contains an even number and its corresponding values h $(*_1)$ and h (GC)) will be, in a safe manner, delivered to users.

Instead of a pair as a public key (2n-1, e), we suggest that the first part of the public key, instead of 2n-1, be h $(*_1)$ and h (GC), based on which the user would set the number 2n by reading the table, and therefore the number of 2n-1 (the RSA modulus) would be known.

It is clear that this procedure does not weaken the RSA. It just makes it difficult for those who intend to reveal the secret key, because prior to the use of algorithms for finding prime factors of the number $2n\text{-}1=p_1*q_1$, that number should be determined primarily, which is very difficult for large numbers if we know only h $(*_1)$ and h(GC).

#### *4.2.2 The Second Possibility*

Another possibility would be publishing the number 2n, which implicitly publishes (GC), too. (GC) may be (another part of the key pairs) public key for RSA (in the standard label e) if gcd((GC),$\theta$)=1, or the first number greater than it that is relatively prime to $\theta$, where $\theta=(p_1\text{-}1)*(q_1\text{-}1)$. This would be a semi-public key cryptosystem, as the users in addition to the secret key d, obtain in the same way, safely, the public key e, while others who have bad intentions must first find (implicitly published) public key e, which is a very demanding job in terms of time, and only then they may access the disclosure of the secret key d. It is clear that in the meantime, we can change the parameters of RSA and thus further complicate efforts to breach confidentiality and integrity of our data.

## 5. CONCLUSIONS

This work, too, is in line with our belief that it is necessary that each country protects the confidentiality and integrity of its data using its own software [7]. Good experts are a prerequisite for this, and they cannot exist without the increased interest of young people in cryptography. We believe that there is no such an interest without more interesting approach to cryptography, and encryption of cryptographic algorithms and experimentation with own software is the best way for that. To this end we have written this paper dealing with such an important topic for practical cryptography: Minimizing the time of detection of large (probably) prime numbers.

The consideration of our problem naturally led to the Goldbach's conjecture [2]. We have noticed that the Goldbach's conjecture can find its place in cryptography because its assumed property can only reduce the detection time (not increase it). It is quite possible that Goldbach's conjecture can play an important role in hindering the intentions of an unauthorized user to find out the secret key mathematically, and thus to compromise the integrity and confidentiality of our data.

Regarded more widely, we believe that the using of Goldbach's conjecture can slow down the trend of massive transition from RSA to ECC. In that case, the increase in the number of bits would no longer be the only asset of RSA.